\documentstyle[12pt]{article}
\begin{document}
\begin{titlepage}
         \title{ Maximal Acceleration Corrections  to \\
                 the Lamb Shift of Hydrogen, \\
                 Deuterium and He$^{+}$}

\author{G. Lambiase$^{a}$\thanks{{\it{E-mail}}: lambiase@vaxsa.csied.unisa.it}
,~~~G.Papini$^{b}$\thanks{{\it{E-mail}}: papini@cas.uregina.ca}~~~and
~~~G. Scarpetta$^{a,c}$ \\
{\em $^a$Dipartimento di Scienze Fisiche ``E.R. Caianiello''}\\ 
{\em  Universit\`a di Salerno, 84081 Baronissi (SA), Italy}\\
{\em  $^a$Istituto Nazionale di Fisica Nucleare, Sezione di Napoli}\\
{\em $^b$Department of Physics, University of Regina,} \\
{\em Regina, Sask. S4S 0A2, Canada}\\
{\em $^c$International Institute for Advanced Scientific Studies} \\ 
{\em Vietri sul Mare (SA), Italy}\\
}
\date{\empty}
\maketitle

              \begin{abstract}
The maximal acceleration corrections to the Lamb shift of one--electron atoms 
are calculated in a non--relativistic approximation. They are compatible with 
experimental results, are in particularly good agreement with 
the $2S-2P$ Lamb shift in hydrogen and
reduce by $\sim 50\%$ the experiment-theory discrepancy for 
the $2S-2P$ shift in He$^+$. 
	      \end{abstract}

\thispagestyle{empty}
\vspace{20. mm}
              \vfill
	      \end{titlepage}

This paper presents the calculation of maximal acceleration (MA)
corrections to the
Lamb shift of one--electron atoms and ions, according to the model of
Caianiello and collaborators \cite{ANI}, \cite{GAS}. 
The view frequently held \cite{TUT}, \cite{BRA} 
that the proper acceleration of a particle is limited upwardly finds in this
model a geometrical interpretation epitomized by the line element
\begin{equation}\label{eq1}
d\tilde{s}^2=\tilde{g}_{\mu\mu}dx^{\mu}dx^{\nu}=
ds^2\left(1-\frac{|\ddot{x}|^2}{{\cal A}_m^2}\right)=
\sigma^2 (x) ds^2
\end{equation}
experienced by the accelerating particle along its worldline. In
(\ref{eq1}) 
${\cal A}_m=2mc^3/\hbar$ is the proper MA of the particle
of mass $m$, $\ddot{x}^{\mu}$ its acceleration  and 
$ds^2=g_{\mu\nu}dx^{\mu}dx^{\nu}$ is the metric due to a background 
gravitational field. In the absence of gravity, $g_{\mu\nu}$ is replaced by the
Minkowski metric tensor $\eta_{\mu\nu}$. Similar results have also been
obtained in the context of Weyl space \cite{WEY} and of a geometrical analogue
of Vigier's stochastic theory \cite{UIG}.

Eq. (\ref{eq1}) has several implications for relativistic kinematics
\cite{SCA}, the energy spectrum of a uniformly accelerated particle 
\cite{CGS}, the periodic structure as a function of momentum in 
neutrino oscillations \cite{CGS},
the Schwarzschild horizon \cite{TTA},  the expansion
of the very early universe \cite{SPE} and the mass of the 
Higgs boson \cite{KUW}.
It also makes the metric observer-dependent, as conjectured by Gibbons
and Hawking \cite{GIB}, and leads in a natural way to hadron 
confinement \cite{PRE}.

The extreme large value that ${\cal A}_m$ takes
for all known particles
(${\cal A}_m\simeq 0.9 \ 10^{30} m$ \ m s$^{-2}$ MeV$^{-1}$)  makes a direct
test of Eq. (\ref{eq1}) very difficult. 
Nonetheless a realistic test that makes use of 
photons in cavities has been recently suggested \cite{INI} and 
attempts in this direction will hopefully lead to conclusive results. 

Recent advances in high resolution spectroscopy are now allowing Lamb shift
mesearements of unprecedented precision, leading in the case of simple atoms and
ions to the most stringent tests of quantum electrodynamics (QED). MA
corrections due to the metric (\ref{eq1}) appear directly in the
Dirac equation for the electron that must now be written in covariant form 
\cite{PAR} and referred to a local Minkowski 
frame by means of the vierbein field
$e_{\mu}^{\,\,\,\, a}(x)$. From (\ref{eq1}) one finds $e_{\mu}^{\,\,\,\, a}
=\sigma(x)\delta_{\mu}^{\,\,\,\, a}$, where Latin indices refer to the locally
inertial frame and Greek indices to a generic non--inertial frame.
The covariant matrices $\gamma^{\mu}(x)$ satisfy the anticommutation 
relations ~$\{\gamma^{\mu}(x), \gamma^{\nu}(x)\}$
$=2\tilde g^{\mu\nu}(x)$, while the  covariant
derivative ${\cal D}_{\mu}\equiv \partial_{\mu}+\omega_{\mu}$ 
contains the total connection $\omega_{\mu}=
\frac{1}{2}\sigma^{ab}\omega_{\mu ab}$, where
$\sigma^{ab}=\frac{1}{4}\,[\gamma^a,\gamma^b]$, 
$\omega_{\mu\,\,\,\,b}^{\,\,\,\,a}=(\Gamma_{\mu\nu}^{\lambda}\,
e_{\lambda}^{\,\,\,\,a}-\partial_{\mu}e_{\nu}^{\,\,\,\,a})
e^{\nu}_{\,\,\,\,b}$
and  $\Gamma_{\mu\nu}^{\lambda}$
represent the usual Christoffel symbols. For conformally flat metrics
$\omega_{\mu}$ takes the form
$\omega_{\mu}=\frac{3}{2\sigma}\sigma^{ab}\eta_{a\mu}\sigma_{,b}$.
By using the transformations
$\gamma^{\mu}(x)=e^{\mu}_{\,\,\,\,a}(x)\gamma^a$
so that $\gamma^{\mu}(x)=\sigma^{-1} (x)\gamma^{\mu}$,
where $\gamma ^{\mu}$ are the usual constant Dirac matrices,
the Dirac equation can be written in the form
\begin{equation}\label{eq3}
\left[ i\hbar\gamma^{\mu}\left(\partial_{\mu}+i\frac{e}{\hbar c}A_{\mu}\right)
+i\frac{3\hbar}{2}\gamma^{\mu}(\ln\sigma)_{,\mu}
-mc\sigma (x)\right]\psi(x)=0\,{.}
\end{equation}
>From (\ref{eq3}) one obtains the Hamiltonian
\begin{equation}\label{eq4}
H= - i\hbar c\vec{\alpha}\cdot \vec{\nabla} + e \gamma^0 \gamma^{\mu}A_{\mu}(x) 
- i\frac{3\hbar c}{2} \gamma^0 \gamma^{\mu}(\ln\sigma )_{,\mu} +
mc^2\sigma(x)\gamma^0\,{,}
\end{equation}
which is in general non--Hermitian \cite{PAR}.
However when one splits the Dirac spinor
into large and small components, the only non-Hermitian
term is $(\ln\sigma )_{,0}$. If $\sigma$
varies slowly in time, or is time-independent, as in the present case,
this term can be neglected and Hermiticity is recovered.

A recent attempt to estimate the Lamb shift corrections due to (\ref{eq1})
was carried out in the local frame of the electron and did not therefore take
into account properly the electromagnetic field experienced by the 
electron \cite{LPS}.
Hamiltonian (\ref{eq4}) corrects this inadequecy. The calculations are
also extended to include the Lamb shift in deuterium and He$^+$.
Here, as in the previous MA calculations\cite{LPS}, the nucleus is
considered to be pointlike and its recoil is neglected.

In QED the Lamb shift corrections are usually calculated by means of 
a non--relativistic approximation \cite{ITZ}. This is also done here.
For the electric field $E(r)=kZe/r^2 (k=1/4\pi\epsilon_{0})$, the 
conformal factor becomes
$ \sigma(r)=(1-\left(\frac{r_0}{r}\right)^4)^{1/2}$,
where 
$ r_0\equiv (kZe^2/m{\cal A}_m)^{1/2}\sim \sqrt{Z}\,2.3\cdot 
10^{-14}\mbox{m} $
and $r>r_0$. The calculation of $\ddot{x}^{\mu}$ is performed classically
in a non--relativistic approximation. This is justified because for 
the electron $v/c$ is at most $\sim 10^{-3}$.
Neglecting contributions of the order
$O({\cal A}_m^{-4})$,  
$ \sigma (r)\sim 1-(1/2)(r_0/r)^4 $.
This expansion requires that in the following only those values of $r$
be chosen that are above a cut--off $\Lambda$, such that for
$r>\Lambda >r_0$ the validity of the expansion is preserved. The actual
value of $\Lambda$ will be selected later.
The length $r_0$ has no fundamental significance in QED and depends in
general on the details of the acceleration mechanism. It is only the distance
at which the electron would attain, classically, the acceleration ${\cal A}_m$
irrespective of the probability of getting there.

By using the expansion for $\sigma (r)$ in (\ref{eq4}) one finds that
all MA effects are contained in the perturbative terms
\begin{equation}\label{eq8}
H_{r_0}=-\frac{mc^2}{2}\left(\frac{r_0}{r}\right)^4\beta+
i\frac{3\hbar c}{4}r_0^4\vec{\alpha}\cdot\vec{\nabla} \frac{1}{r^4}\equiv 
{\cal H}+{\cal H}^{\prime}\,{.}
\end{equation}
By splitting $\psi(x)$ into large
and small components $\varphi$ and $\chi$ and using
$\chi=-i(\hbar/2mc)\vec{\sigma}\cdot\vec{\nabla}\varphi\ll\varphi$
one obtains for the perturbation due to ${\cal H}$ 
\begin{equation}\label{eq9}
\delta {\cal E}_{nlm}\simeq
-\frac{mc^2}{2}r_0^4\int d^3\vec{r}\frac{1}{r^4}
\varphi^{*}_{nlm}\varphi_{nlm}\,{.}
\end{equation}
The perturbation due to ${\cal H}^{\prime}$ vanishes.
In (\ref{eq9}) 
$\varphi_{nlm}$ are 
the well known eigenfunctions for one--electron atoms.
The integrations over the angular variables in (\ref{eq9}) 
can be performed immediately and yield
\begin{eqnarray}
\delta {\cal E}_{20} &=& -\frac{mc^2}{16}\left(\frac{r_0}{a_0}\right)^4
     \left\{\left[4\left(\frac{a_0}{\Lambda}\right)+1\right]e^{-\Lambda/a_0}
-8E_1\left(\frac{\Lambda}{a_0}\right)\right\}\,{,} \label{eq12}\\ 
\delta {\cal E}_{21}&=&-\frac{mc^2}{48}\left(\frac{r_0}{a_0}\right)^4
 e^{-\Lambda/a_0} \,{,}  \label{eq13} \\ 
\delta{\cal E}_{10}&=&-2mc^2\left(\frac{r_0}{a_0}\right)^4\left[
\left(\frac{a_0}{\Lambda}\right)e^{-2\Lambda/a_0}-
            2E_1\left(\frac{2\Lambda}{a_0}\right)\right]\,{,} \label{eq14} 
\end{eqnarray}
where $ E_1 (x)=\int_{1}^{\infty} dy\,e^{-xy}/y$ and
$a_0$ is the Bohr radius divided by $Z$.
In order to calculate the $2S-2P$ Lamb shift corrections it is now necessary 
to choose the value
of the cut--off $\Lambda$. While in QED Lamb shift and fine structure effects are 
cut--off independent, the values of the corresponding MA
corrections increase when $\Lambda$ decreases. This can be understood 
intuitively because the electron finds itself in regions of higher electric 
field at smaller values of $r$. $\Lambda$ is a characteristic
length of the system. It must also represent a distance from the nucleus that
can be reached by the electron whose acceleration and relative perturbations
depend on the position attained.
One may tentatively choose $\Lambda\sim a_0$. 
According to the wave functions involved, the probability that the electron be
at this distance ranges between $0.1$ and $0.5$. Smaller values of $\Lambda$
lead to larger acceleration corrections, but are reached with much lower
probabilities. This is the case of the Compton wavelength of the electron whose use as a cut--off is therefore ruled out in the present context.
For $\Lambda\sim a_0$, 
Eqs. (\ref{eq12})--(\ref{eq14}) give the corrections to the 
levels $2S, 2P$ and $1S$ ($Z=1$)
$\delta {\cal E}_{20}\sim -22.96\,\mbox{kHz}$,
$\delta {\cal E}_{21}\sim -33.42\,\mbox{kHz}$,
$\delta {\cal E}_{10}\sim -325.45\, \mbox{kHz}$,
yielding the Lamb shift correction
$\delta {\cal E}_L=\delta {\cal E}_{20}-\delta {\cal E}_{21}
\sim +10.46\,\mbox{kHz}$.
The results are summarized below \cite{ACU}.

{\it a) $2S-2P$ Lamb shift for Hydrogen}. The most recent experimental and 
theoretical values of the classic Lamb shift are reported in Table I
and compared with the theory with MA corrections. These amount to
$\delta {\cal E}_L$ above.
$r_p$ is the rms charge radius of the proton \cite{RP1}, \cite{RP2}. 
The MA corrections are in very good agreement with all experimental results 
reported and the value $r_p=0.862$fm. They also appear to be consistently in 
the right direction.
The coefficients of (\ref{eq12})--(\ref{eq14})
are proportional to powers $(Z\alpha)^6$ from which it 
follows that the MA corrections are comparable in magnitude with
those obtained from perturbative QED up to order $\alpha^7$.
Further improvements in experimental sensitivity 
might indeed be able to distinguish between the MA and QED contributions.
Unfortunately, higher experimental precision seems difficult to achieve because
of the $100\mbox{MHz}$ natural linewidth of the $2P$ state. 

{\it b) $1S$ ground state Lamb shift $L_{1S}$ in Hydrogen}.
Mesearements of $L_{1S}$ have recently become very precise by 
comparison of the $1S-2S$ resonance with four times the 
frequencies of the $2S-4S$ and $2S-4D$
two-photon transitions. The MA corrections are given 
by $\delta {\cal E}_{10}$ above. 
The results are compared in Table II.
The first line repeats the
results before 1992. Experiment and theory were known to agree (within
$0.1\mbox{MHz}$) for $r_p = 0.805\mbox{fm}$. The MA corrections also agree
within $0.2\mbox{MHz}$.
More recently a discrepancy has appeared between experiment and theory with the
adoption of the more reliable value $r_p = 0.862\mbox{fm}$ 
increasing the theoretical estimate to ~$8173.12(6)\mbox{MHz}$. The
agreement is improved in this instance by the MA corrections for the choice of
the new radius. More recent experimental and theoretical data are
compared on the last three lines of Table II.  
The MA corrections would restore by themselves the agreement between
experiment and QED without two loop corrections. However the agreement 
between experiment and QED improves significantly when the 
two-loop corrections
calculated by Pachucki \cite{PAC} are included in the theoretical estimate and
the MA corrections are excluded. These latter effects shift the theoretical
estimate by $\sim 0.3\mbox{MHz}$ below the experimental results. It is
interesting to observe that the dominant MA correction, of order $\alpha^6$,
is approximately of the same magnitude of the two-loop correction of order
$\alpha^7$ which must therefore be considered as truly large. While the 
Pachucki calculation restores the agreement between theory and 
experiment for  hydrogen, it upsets that of 
the $2S-2P$ splitting of $He^+$ \cite{NEW}. 

The MA contributions (\ref{eq12})-(\ref{eq14}) are particularly
sensitive to the choice of $\Lambda$. For instance, a $10\%$ increase in the
value of $\Lambda$ shifts upward the MA correction 
from $-325\mbox{kHz}$ to $-230\mbox{kHz}$, and improves the agreement
between experiment and MA theory considerably. This is
largely due to the presence of a ground state wave function peak at $r=a_0$. 

{\it c) Lamb shift $\frac{1}{4}L_{1S} - \frac{5}{4}L_{2S}+L_{4S}$ in Hydrogen
and Deuterium}. These are mesearements of the $L_{1S}$ Lamb shift by direct 
comparison of the $1S-2S$ with the $2S-4S$ two-photon transitions. The MA
corrections are determined from (\ref{eq9}) and the
corresponding hydrogenic eigeinfunctions and are 
$L_{4S}=-2.54\mbox{kHz}$
and ~$\frac{1}{4}L_{1S} - \frac{5}{4}L_{2S}+L_{4S}=-55\mbox{kHz}$.
The results are compared in Table III, where $r_{ch}$ is the rms charge radius
of the nucleus.
The agreement between experiment and theory is good for hydrogen and remains
good with the introduction of MA.
For deuterium the agreement is still reasonable because of uncertainties in the
measurement of charge and matter radii. The introduction of MA lowers
theoretical estimates by $55\mbox{KHz}$, which is in the right direction.
The MA estimate based on the earlier calculation \cite{EXB} still falls within
the experimental error of the most recent measurement \cite{NEW}.

{\it d) $L_{1S}$ for Deuterium}. The MA correction is $\delta E_{10}$
and the results are summarized in Table IV.
The agreement of the MA theory with experiment is again better
in the absence of two-loop corrections. When these are included, the theory
falls short by approximately $270\mbox{kHz}$.

{\it e) Lamb shift $2S-2P$ for $He^+$}. The MA corrections is 
here ~$+0.527\mbox{MHz}$. The results are given in Table V.
While the agreement between experiment \cite{AVV} and theory was 
good ~($\sim 10\mbox{kHz}$) before the introduction
of two-loop corrections, the latter have
introduced a discrepancy of $\sim 1.27\mbox{MHz}$ \cite{NEW} to $\sim 1.190\mbox{MHz}$ \cite{AV1}. The method of measurement used to obtain the $He^+$ result \cite{AVV} has been recently verified by a parallel high-precision measurement of the Lamb shift in $H$ \cite{AV1}. The discrepancy must be treated seriously and is unresolved.The MA contributions reduce
significantly the disagreement with theory to $\sim 0.74\mbox{MHz}$ and 
$\sim 0.66\mbox{MHz}$, respectively. 
If the $He^+$ experiments will confirm the predictions of QED, then the Lamb
shift measurements in hydrogen will determine the proton radius to within a
few percent \cite{BHB}.

In conclusion, the agreement between MA corrections and experiment is at
present very good for the $2S-2P$ Lamb shift in hydrogen ($\sim 7\mbox{kHz}$)
and comparable with the agreement of experiments with standard QED 
with and without two-loop corrections.
The agreement is also good for the $\frac{1}{4}L_{1S}-\frac{5}{4}L_{2S}+L_{4S}$
in Lamb shift in hydrogen and comparable, in some instances, with that between
experiment and QED ($\sim 30 \mbox{kHz}$). The
corresponding MA corrections for deuterium fare worse than the conventional
theory, but no worse than the disagreement ($\sim 38\mbox{kHz}$) between the
two QED estimates considered. For the $L_{1S}$ case in deuterium,
the MA theory is worse ($\sim -270\mbox{kHz}$) than the standard
one in reproducing the experimental data when two-loop corrections are included,
but better than QED alone when these are excluded. The latter
statement also applies to the $L_{1S}$ shift in hydrogen. 
Finally, the MA corrections improve the agreement between experiment and 
theory by $\sim 50\%$ for the $2S-2P$
shift in $He^+$. While the two-loop corrections have been independently
confirmed by two groups \cite{BHB}, there seems room for improvement on the
experimental side regarding the sizes of proton and of deuterium and the 
nuclei. At the same time new experiments, now in planning stages \cite{GIO},
should resolve some of the discrepancies now existing between experiment
and QED and ultimately provide stringent tests of the MA theory.

\bigskip
\bigskip

Research supported by MURST fund $40\%$ and $60\%$, DPR 382/80, the Natural
Sciences and Engineering Research Council of Canada and NATO Collaborative 
Research Grant No. 970150.
G.P. gladly acknowledges the continued research support 
of Dr. K. Denford, Dean of Science and Dr. L. Symes V. President 
Research, University of Regina.
G.L. wishes to thank  Dr. K. Denford for his kind hospitality 
during a stay at the University of Regina and V.V. Nesterenko
for useful discussions. 

\newpage

\begin{center}
\centering{Table I. $2S-2P$ Lamb shift for Hydrogen}\\
\end{center}
\begin{center}
\begin{tabular}{cccc}\hline
{\it Experiment} & {\it Theory} & $r_p$  & MA \\ 
 ($\mbox{kHz}$)    & ($\mbox{kHz}$)    &  (fm)  & ($\mbox{kHz}$) \\ \hline
1057845(9)\cite{LUN} & $1057810(4)(4)^a$\cite{EID} & 0.805(11) & 1057820.46\\ 
1057851.4(19)\cite{PAL} & $1057829(4)(4)^a$\cite{EID} & 0.862(12) 
                                                     & 1057839.46\\
1057839(12)\cite{HAG} & $1057838(6)^b$\cite{PAC} & 0.862(12) & 1057848.46 \\
$1057842(12)^c$\cite{JPK} & 1057839(4)\cite{GIO} & 0.862(12) & 1057849.46 \\ 
                                                                \hline
\end{tabular}
\end{center}            
\small{a: correction to order $\alpha^2(Z\alpha )^5 m$ \\
       b: two loop corrections, $\alpha^2(Z\alpha )^5 m$ \\
       c: result of Ref. \cite{HAG} amended to take into account a new value
          of $\alpha$} 

\newpage

\begin{center}
\centering{Table II. $L_{1S}$ in Hydrogen}
\end{center}
\begin{center}
\begin{tabular}{cccc}\hline
{\em Experiment}      &  {\em Theory}        & $r_p$ &  MA       \\ 
 ($\mbox{MHz}$)       &  ($\mbox{MHz}$)       & (fm)  & ($\mbox{MHz}$) \\ \hline
8172.82(11)\cite{EXA} & 8172.94(9)\cite{EXA} & 0.805       & 8172.615  \\ 
8172.86(6)\cite{EXB}  & 8172.97\cite{EXB}    & 0.805       & 8172.645  \\
                      & 8172.654(40)\cite{BHB} & 0.805     & 8172.329  \\
                      & 8173.12(6)\cite{EXB} & 0.862       & 8172.795  \\ 
8172.874(60)\cite{NEW} & 8173.097(40)$^a$\cite{NEW} & 0.862 & 8172.772 \\
                       & 8172.802(40)$^b$\cite{PAC}  & 0.862 & 8172.477 \\
8172.827(51)\cite{BHB} & 8172.802(30)$^b$\cite{BHB}  & 0.862 & 8172.477 \\
\hline
\end{tabular}
\end{center}
{\small {a: without two-loop corrections \\
         b: with two-loop corrections   }}

\newpage

\begin{center}
\centering{Table III. $\frac{1}{4}L_{1S} - \frac{5}{4}L_{2S}+L_{4S}$}
\end{center}
\begin{center}
\begin{tabular}{ccccc}\hline
{\em Experiment}   &   {\it Theory} &   & $r_{ch}$     & MA \\ 
 ($\mbox{MHz}$)     &  ($\mbox{MHz}$)  &  & (fm) & ($\mbox{MHz}$) \\
\hline
                                 &       & Hydrogen &          &       \\
868.61(3)\cite{EXA}              & 868.64(2)  &  & 0.805     &  868.585 \\
                                 & 868.66(2)  &  & 0.862     &  868.605 \\
868.630(12)\cite{PAC},\cite{EXB} & 868.623(5) &  & 0.862     &  868.568 \\
                                 & 868.656$ $ &  & 0.862     &  868.601  \\
                      &                 & Deuterium &              &        \\
869.839(21)\cite{EXB} & 869.8624\cite{EXB} & & 2.115(6)\cite{EXB} & 869.807  \\
869.826(20)\cite{NEW} & 869.8243\cite{NEW} & & 2.115(6)\cite{NEW} & 869.769 \\
\hline
\end{tabular}
\end{center}

\newpage

\begin{center}
\centering{Table IV. $L_{1S}$ for Deuterium.}
\end{center}
\begin{center}
\begin{tabular}{cccc}\hline
{\it Experiment}  &  {\it Theory}   &  $r_{ch}$  &  MA  \\
  ($\mbox{MHz}$)  &  ($\mbox{MHz}$) &  (fm)      &  ($\mbox{MHz}$) \\ \hline
8184.00(8)\cite{EXB} & 8184.13(6)\cite{EXB} & 2.115 & 8183.805 \\
8183.807(78)\cite{NEW} & 8184.080(47)$^a$\cite{NEW} & 2.115 & 8183.755 \\
                       & 8183.785(47)$^b$\cite{NEW} &    &  8183.460 \\ \hline
\end{tabular}
\end{center}
{\small a: without two-loop corrections \\
        b: with two-loop corrections }

\newpage

\begin{center}
\centering{Table V. $2S-2P$ Lamb shift for $He^+$}
\end{center}
\begin{center}
\begin{tabular}{ccc}\hline
{\it Experiment}  &  {\it Theory}   &   MA  \\
($\mbox{MHz}$)    &  ($\mbox{MHz}$) &  ($\mbox{MHz}$) \\ \hline
14042.52(16)\cite{AVV} & 14042.51(20)$^a$\cite{AVV} & 14043.037       \\
                       & 14041.25$^b$\cite{NEW} & 14041.777      \\ 
                       & 14041.33$^b$\cite{AV1} & 14041.857      \\ \hline
\end{tabular}
\end{center}
{\small a: without two-loop corrections \\
        b: with two-loop corrections }


\begin{thebibliography}{99}
\bibitem{ACU}  
    {\small{ It has also been suggested that the MA be referred 
    to Planck's mass \cite{BRA}. Then ${\cal A}_P=m_Pc^3/\hbar\sim 5\cdot
    10^{51}m/sec^2$ and $\Lambda \sim a_0$ gives $\delta {\cal E}_L\sim
    7.8\cdot 10^{-41}$Hz. A cut-off that yields seemingly reasonable values of
    $\delta {\cal E}_L$ may be derived from the requirement that in the
    expansion of $\sigma (r)$ terms $(r_0/r)^4$ be at least $\sim 10^{-2}$.
    This leads to $\Lambda\sim 100^{1/4}r_0\sim 4.8\cdot 10^{-25}$m
    and $\delta {\cal E}_L\sim 1.6$Hz, which must be construed as an upper
    limit. However, the probability that the electron be so close to the origin
    (indeed well inside the nucleus) is vanishingly small.}} 
\normalsize
\end{thebibliography}
\end{document}